# On the High-SNR Capacity of Non-Coherent Networks

Amos Lapidoth*

**Abstract**

We obtain the first term in the high signal-to-noise ratio (SNR) expansion of the capacity of fading networks where the transmitters and receivers—while fully cognizant of the fading *law*—have no access to the fading *realization*. This term is an integer multiple of log log SNR with the coefficient having a simple combinatorial characterization.

## 1 Introduction

In this paper we consider a discrete-time vector fading channel, where the transmitted vector suffers from both multiplicative and additive noises. The multiplicative noise takes the form of a matrix-valued stationary and ergodic process that multiplies the transmitted vector, and the additive noise takes the form of independent and identically distributed (IID) isotropic Gaussian vectors. We only consider the case where neither the realization of the additive noise nor of the multiplicative noise is known to the transmitter and receiver; only their probability laws are given. The mathematical model that we address is thus very similar to the "non-coherent" flat-fading multiple-antenna channel model.

There is, however, an important difference. In the multiple-antenna channel model we think of the components of the transmitted vector as being the signals transmitted by co-located antennas. Similarly, the components of the received vectors are viewed as the signals received at co-located antennas. Our model is more general. We can think of the different components of the input vector as being controlled by a single-user as in a single-user multiple-antenna communication scenario, but we can also think of each component as being controlled by different geographically separated users as, for example, in a multiple-user network where each of the users employs a single transmit antenna. We can also envision that the components of the transmitted vector are partitioned into disjoint groups where the different groups are controlled by geographically separated users. This corresponds to a network where the different geographically separated users may employ multiple transmit antennas of various numbers. Finally, in our setup the different components of the input vector need not correspond to physically different transmit antennas. We can also envision a scenario where they correspond to transmissions taking place at different frequencies and/or times as in a network employing a slotted protocol. Analogous scenarios can be envisioned for the received vector.

The various scenarios mentioned above differ not only in the allowed dependencies between the different components of the transmitted vector. It turns out that, at high signal-to-noise ratio (SNR), far more important is the structure of the multiplicative noise that they imply. For example, if a certain receive antenna and a certain transmit antenna operate at different time/frequency slots, then the corresponding component in the multiplicative noise matrix will be deterministically zero. A similar situation occurs when a given transmitter is geographically very far apart from a given receiver as could, for example, be the case in a cellular system. For example, in Wyner's linear cellular model [1] [2] the transmitters are assumed to be uniformly spaced on a line, and each transmitter is received by only two base-stations: the base station to its left and the base station to its right.

As we shall see, rather than the cooperation restrictions, it is these deterministic zeroes that will determine the high SNR asymptotic behavior of channel capacity. Very roughly, the main result of this paper is that, irrespective of the cooperation allowed, at high SNR the capacity of the channel $C$ is given approximately by

$$C \approx \kappa^* \cdot \log\log \text{SNR} \tag{1}$$

where the non-negative integer $\kappa^*$ can be computed combinatorially from the zeros of the multiplicative noise.

The above result can be viewed as an extension of a result of [3] on multiple-antenna fading channels. In the multiple-antenna scenario where the components of the transmitted vector are geographically co-located and where the components of the received vector are also co-located, there are typically no deterministic zeroes in the fading noise. In this case it can be readily verified that our combinatorial expression for $\kappa^*$ yields the value of 1, thus recovering the $1 \cdot \log\log \text{SNR}$ asymptotics of [3].

The rest of the this paper is organized as follows. In the next section we describe the channel model and state the main result. In Section 3 we provide a proof, and in the final section, Section 4, we summarize our results and discuss some possible extensions.

## 2 Channel Model and Main Result

The channel we consider is a discrete-time channel where the time-$k$ channel input $\mathbf{x}_k \in \mathbb{C}^{n_\text{T}}$ is an $n_\text{T}$-dimensional complex vector, where $k \in \mathbb{Z}$ is a discrete-time index taking value in the integers $\mathbb{Z}$; $n_\text{T}$ is a positive integer; $\mathbb{C}$ denotes the complex field; and $\mathbb{C}^{n_\text{T}}$ denotes the $n_\text{T}$-dimensional complex Euclidean space. We refer to $n_\text{T}$ as the number of transmitters, and to the set

$$\mathcal{T} = \{1, \ldots, n_\text{T}\} \tag{2}$$

as the set of transmitters. For every $t \in \mathcal{T}$ we denote the $t$-th component of the time-$k$ input vector $\mathbf{x}_k$ by $x_k(t)$. This corresponds to the signal transmitted at time $k$ by Transmitter $t$. The time-$k$ channel output $\mathbf{Y}_k \in \mathbb{C}^{n_\mathrm{R}}$ corresponding to the input $\mathbf{x}_k$ is given by

$$\mathbf{Y}_k = \mathbb{H}_k \mathbf{x}_k + \mathbf{Z}_k \tag{3}$$

where $n_\mathrm{R}$ is a positive integer that denotes the number of receive antennas and where

$$\mathcal{R} = \{1, \ldots, n_\mathrm{R}\} \tag{4}$$

denotes the set of receivers. In the above, $\{\mathbb{H}_k\}$ is a matrix-valued stochastic process such that at every time instant $k$ the random matrix $\mathbb{H}_k$ is an $n_\mathrm{R} \times n_\mathrm{T}$ complex random matrix, and the random vectors $\{\mathbf{Z}_k\}$ are independent and identically distributed (IID), each taking value in $\mathbb{C}^{n_\mathrm{R}}$ according to an isotropic circularly symmetric multivariate complex Gaussian law

$$\mathbf{Z}_k \sim \mathcal{N}_\mathbb{C}(\mathbf{0}, \mathsf{I}_{n_\mathrm{R}}) \tag{5}$$

where $\mathsf{I}_{n_\mathrm{R}}$ denotes the $n_\mathrm{R} \times n_\mathrm{R}$ identity matrix. (In general, $\mathbf{W} \sim \mathcal{N}_\mathbb{C}(\boldsymbol{\mu}, \Lambda)$ indicates that $\mathbf{W} - \boldsymbol{\mu}$ is a zero-mean circularly symmetric complex Gaussian random vector of covariance matrix $\Lambda$.) We assume throughout that the processes $\{\mathbb{H}_k\}$ and $\{\mathbf{Z}_k\}$ are independent and that their joint law does not depend on the input sequence $\{\mathbf{x}_k\}$. Denoting by $H_k(r, t)$ the row-$r$ column-$t$ entry of the matrix $\mathbb{H}_k$, and denoting by $Z_k(r)$ the $r$-th component of the time-$k$ additive noise vector $\mathbf{Z}_k$, we can re-write (3) as

$$Y_k(r) = \sum_{t \in \mathcal{T}} H_k(r, t) x_k(t) + Z_k(r), \qquad r \in \mathcal{R}. \tag{6}$$

To account for the possibility that some of the components of the fading matrices might be deterministically zero we introduce the set $\mathcal{Z}$

$$\mathcal{Z} \subset \mathcal{R} \times \mathcal{T} \tag{7}$$

where if $(r, t) \in \mathcal{Z}$ then $H_k(r, t)$ is deterministically zero at all times $k \in \mathbb{Z}$:

$$(r, t) \in \mathcal{Z} \Rightarrow \Big(H_k(r, t) = 0, \quad \forall k \in \mathbb{Z}\Big). \tag{8}$$

As for the other components, we shall assume a finite second moment

$$\mathsf{E}\big[|H_k(r, t)|^2\big] < \infty, \qquad (r, t) \in \mathcal{R} \times \mathcal{T} \tag{9}$$

and a finite differential entropy rate condition that we next describe. But first we introduce some notation. Given a collection of random variables $\{W(\alpha)\}_{\alpha \in \mathcal{A}}$ indexed by a set $\mathcal{A}$ we denote, for any subset $\mathcal{B} \subseteq \mathcal{A}$, by $W(\mathcal{B})$ the unordered collection $\{W(\alpha)\}_{\alpha \in \mathcal{B}}$. With this notation and (7) we have that $H_k(\mathcal{Z}^\mathrm{c})$ is the collection of $|\mathcal{Z}^\mathrm{c}|$ $(= n_\mathrm{R} \cdot n_\mathrm{T} - |\mathcal{Z}|)$ random variables

$$H_k(\mathcal{Z}^\mathrm{c}) = \{H_k(r, t) : (r, t) \notin \mathcal{Z}\} \tag{10}$$

where we use $\mathcal{Z}^\mathrm{c}$ to denote the set complement of $\mathcal{Z}$ in $\mathcal{R} \times \mathcal{T}$ and we use $|\cdot|$ to denote set cardinality. The finite differentiable entropy rate condition that we require can be now stated as

$$h\big(\{H_k(\mathcal{Z}^\mathrm{c})\}_{k \in \mathbb{Z}}\big) > -\infty. \tag{11}$$

In the case where $\{\mathbb{H}_k\}$ is IID, this condition translates to the joint differential entropy of the $(n_\mathrm{R} \cdot n_\mathrm{T} - |\mathcal{Z}|)$ random variables $\{H_k(r,t), \quad (r,t) \notin \mathcal{Z}\}$ being finite. In the more general case, (11) can be written as

$$\lim_{n\to\infty} \frac{1}{n} h\Big(H_1(\mathcal{Z}^c), \ldots, H_n(\mathcal{Z}^c)\Big) > -\infty \tag{12}$$

or even more explicitly as

$$\lim_{n\to\infty} \frac{1}{n} h\Big(\{H_k(r,t)\}, \quad 1 \leq k \leq n, \quad (r,t) \notin \mathcal{Z}\Big) > -\infty. \tag{13}$$

Notice that a stationary process $\{\mathbb{H}_k\}$ simultaneously satisfies (9) and (11) if, and only if, it simultaneously satisfies (9) and the two conditions

$$h\left(\mathbb{H}_1(\mathcal{Z}^c)\right) > -\infty \quad \text{and} \quad \lim_{k\to\infty} I\left(\mathbb{H}_1, \ldots, \mathbb{H}_{k-1}; \mathbb{H}_k\right) < \infty. \tag{14}$$

We denote by $C_\mathrm{SU}(\mathcal{E})$ the capacity of this channel under full cooperation conditions when the input is subjected to the average power constraint $\mathcal{E}$. That is,

$$C_\mathrm{SU}(\mathcal{E}) = \lim_{n\to\infty} \frac{1}{n} \sup I(\mathbf{X}_1, \ldots, \mathbf{X}_n; \mathbf{Y}_1, \ldots, \mathbf{Y}_n) \tag{15}$$

where the supremum is over all joint distributions on $\mathbf{X}_1, \ldots, \mathbf{X}_n$ satisfying

$$\frac{1}{n} \sum_{k=1}^n \mathsf{E}\big[\|\mathbf{X}_k\|^2\big] \leq \mathcal{E}.$$

This is thus the capacity when a single-user controls the input vector $\mathbf{x}_k \in \mathbb{C}^{n_\mathrm{T}}$, and when a "super-receiver" has access to all the components of the received vector $\mathbf{Y}_k$. Similarly, we define $C_{\mathrm{SU,FB}}(\mathcal{E})$ as the single-user capacity but when there is a noiseless feedback link so that the time-$k$ transmitted signal $\mathbf{X}_k$ is allowed to depend not only on the message to be transmitted but also on all the past channel outputs. Clearly, $C_\mathrm{SU}(\mathcal{E}) \leq C_{\mathrm{SU,FB}}(\mathcal{E})$ because the feedback link can always be ignored.

At the other extreme we define $C_\mathrm{MAC}(\mathcal{E})$ as the sum-rate capacity for this channel when it is viewed as a multiple-access channel (MAC) where the different components of the input vector are viewed as separate users who wish to communicate independent messages. Each user is assumed to be allowed a peak power of $\mathcal{E}$. The assumption of a "super-receiver" continues to hold. (We shall later see that this assumption can be significantly relaxed.) We assume no feedback link. We thus have,

$$C_\mathrm{MAC}(\mathcal{E}) \leq C_{\mathrm{SU,FB}}(n_\mathrm{T} \cdot \mathcal{E}). \tag{16}$$

To state the paper's main theorem we need to introduce the notion of a "power chain". To define this concept we introduce the following notation. For any transmitter $t \in \mathcal{T}$ let $\mathcal{R}_t$ be the set of receivers that can "hear" it, i.e.,

$$\mathcal{R}_t = \big\{r \in \mathcal{R} : (r,t) \notin \mathcal{Z}\big\}. \tag{17}$$

Analogously, for any receiver $r \in \mathcal{R}$, let $\mathcal{T}_r$ denote the set of transmitters that $r$ "hears":

$$\mathcal{T}_r = \big\{t \in \mathcal{T} : (r,t) \notin \mathcal{Z}\big\}. \tag{18}$$

**Definition 1.** *We shall say that the $\kappa$-tuple $(t_1, \ldots, t_\kappa) \in \mathcal{T}^\kappa$ is a $\kappa$-length power chain with respect to the set $\mathcal{Z}$ if*

$$\mathcal{R}_{t_1} \neq \emptyset \tag{19}$$

*and*

$$\mathcal{R}_{t_\nu} \setminus \bigcup_{1 \leq \eta < \nu} \mathcal{R}_{t_\eta} \neq \emptyset, \quad \nu = 2, \ldots, \kappa. \tag{20}$$

We can now state the paper's main result.

**Theorem 2.** *Consider a vector fading channel (3) whose input takes value in $\mathbb{C}^{n_T}$ and whose output takes value in $\mathbb{C}^{n_R}$. Let the set $\mathcal{Z} \subset \mathcal{R} \times \mathcal{T}$ be given, where $\mathcal{R}$ and $\mathcal{T}$ are defined in (4) and (2) respectively. Assume that the stationary and ergodic matrix-valued fading process $\{\mathbb{H}_k\}$ satisfies (8), (9), and (11). Further assume that $\{\mathbf{Z}_k\}$ are IID according to (5), that the process $\{\mathbf{Z}_k\}$ is independent of $\{\mathbb{H}_k\}$, and that their joint law does not depend on the channel input sequence $\{\mathbf{x}_k\}$. Let $C_{\mathrm{SU,FB}}(\mathcal{E})$ and $C_{\mathrm{MAC}}(\mathcal{E})$ be defined as above. Then,*

$$\overline{\lim_{\mathcal{E} \to \infty}} \left\{ C_{\mathrm{SU,FB}}(\mathcal{E}) - \kappa^* \log \log \mathcal{E} \right\} < \infty \tag{21}$$

*where $\kappa^* = \kappa^*(n_T, n_R, \mathcal{Z})$ is the length of the longest power chain with respect to $\mathcal{Z}$.*

*If, additionally, $\{\mathbb{H}_k\}$ has a Gaussian marginal, i.e., if the components of the matrix $\mathbb{H}_1$ (and hence, by stationarity, of $\mathbb{H}_k$ for any $k$) are jointly circularly symmetric and Gaussian, then*

$$\overline{\lim_{\mathcal{E} \to \infty}} \left\{ \kappa^* \log \log \mathcal{E} - C_{\mathrm{MAC}}(\mathcal{E}) \right\} < \infty. \tag{22}$$

*Moreover, in this Gaussian case, (22) is achievable with $\kappa^*$ single-user scalar detectors. That is, there exist transmitters $t_1, \ldots, t_{\kappa^*} \in \mathcal{T}$; receivers $r_1, \ldots, r_{\kappa^*} \in \mathcal{R}$; and distributions for $\mathbf{X}$ under which the components of $\mathbf{X}$ are independent, under which the peak constraints $|X(t)| \leq \mathcal{E}$, $t \in \mathcal{T}$ are satisfied almost surely, and such that*

$$\overline{\lim_{\mathcal{E} \to \infty}} \left\{ \kappa^* \log \log \mathcal{E} - \sum_{\nu=1}^{\kappa^*} I\big(X(t_\nu); Y(r_\nu)\big) \right\} < \infty. \tag{23}$$

Note that since $\log \log(a\xi) - \log \log \xi$ converges to zero as $\xi \to \infty$ with $a > 0$ held fixed, it follows from (16) and from the theorem that

$$\lim_{\mathcal{E} \to \infty} \left\{ C_{\mathrm{SU,FB}}(\mathcal{E}) - C_{\mathrm{MAC}}(\mathcal{E}) \right\} \geq 0. \tag{24}$$

Consequently, we can loosely say that, at high SNR, the capacity of a Gaussian fading network is given by (1), where $\kappa^* = \kappa^*(n_T, n_R, \mathcal{Z})$, irrespective of whether we impose individual peak power constraints or whether we impose combined average power constraints, irrespective of whether we allow cooperation between the transmitters or not, and irrespective of whether feedback is available or not.

# 3 Proof of Theorem 2

In this section we provide a proof of Theorem 2. We shall begin by showing that it suffices to prove the theorem in the case where the fading $\{\mathbb{H}_k\}$ is memoryless, *i.e.*, when $\{\mathbb{H}_k\}$ are IID. We shall then separately prove the "converse" (21) and the "direct" part (22) in the two corresponding subsections.

Let then $\{\mathbb{H}_k\}$ be some stationary and ergodic fading process with memory satisfying (9) & (14), and let $\{\tilde{\mathbb{H}}_k\}$ be an IID fading process of equal marginal so that the law of $\tilde{\mathbb{H}}_k$ is the same as the law of $\mathbb{H}_1$ (which is the same, by stationary, as the law of $\mathbb{H}_k$ for any $k \in \mathbb{Z}$).

That it suffices to prove the converse in the memoryless case follows because, as shown by Moser [4, Chapter 8], the difference between the feedback-capacity of the channel with fading $\{\mathbb{H}_k\}$ and the capacity of the channel with IID fading $\{\tilde{\mathbb{H}}_k\}$ is bounded in the SNR. For the sake of completeness we repeat Moser's result in Appendix A.

As to the direct part, we note that the capacity of the channel with fading $\{\mathbb{H}_k\}$ cannot be smaller than that of fading $\{\tilde{\mathbb{H}}_k\}$. Indeed, if $Q$ is any distribution on $\mathbb{C}^{n_T}$ then the mutual information on the memoryless channel of fading $\{\tilde{\mathbb{H}}_k\}$ is achievable on the channel of fading $\{\mathbb{H}_k\}$ by considering inputs $\mathbf{X}_1, \ldots, \mathbf{X}_n$ that are IID according to $Q$. Indeed, for such IID inputs

$$\frac{1}{n} I(\mathbf{X}_1^n; \mathbf{Y}_1^n) = \frac{1}{n} \sum_{k=1}^{n} I(\mathbf{X}_k; \mathbf{Y}_1^n | \mathbf{X}_1^{k-1})$$
$$= \frac{1}{n} \sum_{k=1}^{n} I(\mathbf{X}_k; \mathbf{Y}_1^n, \mathbf{X}_1^{k-1})$$
$$\geq \frac{1}{n} \sum_{k=1}^{n} I(\mathbf{X}_k; \mathbf{Y}_k)$$
$$= I(\mathbf{X}_1; \mathbb{H}_1 \mathbf{X}_1 + \mathbf{Z}_1)$$
$$= I\left(\mathbf{X}_1; \tilde{\mathbb{H}}_1 \mathbf{X}_1 + \mathbf{Z}_1\right)$$

where the first equality follows from the chain rule; the subsequent from the independence of $\mathbf{X}_1, \ldots, \mathbf{X}_n$; and the subsequent inequality because reducing observations cannot increase mutual information. Here we use $\mathbf{X}_\ell^m$ to denote $\mathbf{X}_\ell, \ldots, \mathbf{X}_m$.

We shall thus proceed to prove the theorem assuming that the fading is memoryless. In this case we shall omit the time index so that our assumptions on the fading process can be now written as:

$$\mathsf{E}\big[|H(r,t)|^2\big] < \infty, \qquad (r,t) \in \mathcal{R} \times \mathcal{T} \tag{25}$$

$$h\big(H(\mathcal{Z}^c)\big) > -\infty \tag{26}$$

$$(r,t) \in \mathcal{Z} \Rightarrow \Big(H(r,t) = 0, \quad \text{almost surely}\Big). \tag{27}$$

We shall further assume that none of the rows of $\mathbb{H}$ is deterministically zero, *i.e.*,

$$\forall r \in \mathcal{R} \quad \exists t \in \mathcal{T} : (r,t) \notin \mathcal{Z} \tag{28}$$

or equivalently,
$$\mathcal{T}_r \neq \emptyset, \quad r \in \mathcal{R}. \tag{29}$$

This corresponds to the condition that every receiver "hears" at least one transmitter. Analogously, we shall assume that none of the columns of $\mathbb{H}$ is deterministically zero, i.e.,
$$\forall t \in \mathcal{T} \quad \exists r \in \mathcal{R} : (r, t) \notin \mathcal{Z} \tag{30}$$

or equivalently
$$\mathcal{R}_t \neq \emptyset, \quad t \in \mathcal{T}. \tag{31}$$

This corresponds to the condition that every transmitter is heard by at least one receiver. The above assumptions can be made without loss of generality because a receiver that hears no signals (other than ambient additive noise) does not affect the longest power chain and can also be ignored at the detector. Similarly, a transmitter that cannot be heard by any receiver will never be an element of a power chain and there is also no point in having it transmit any signal.

## 3.1 The Converse

In this section we provide a proof of (21) for IID fading $\{\mathbb{H}_k\}$ satisfying (25), (26), and (27). We begin by considering the "ordering permutation" $\sigma(\mathbf{x})$ of a given $n_\text{T}$-tuple $\mathbf{x} \in \mathbb{C}^{n_\text{T}}$. This is the permutation that orders the components of $\mathbf{x}$ in descending order of their magnitudes, resolving ties with preference to lower indices. Thus, given an $n_\text{T}$-tuple $\mathbf{x} \in \mathbb{C}^{n_\text{T}}$ we set $\sigma(\mathbf{X})$ to be the permutation $\tau : \nu \mapsto \tau_\nu$ on $\mathcal{T}$ that satisfies
$$|x(\tau_1)| \geq |x(\tau_2)| \geq \cdots \geq |x(\tau_{n_\text{T}})| \tag{32}$$

and that resolves ties in favor of lower indices so that
$$|x(\tau_\nu)| = |x(\tau_{\nu+1})| \Rightarrow \tau_\nu < \tau_{\nu+1}. \tag{33}$$

The form in which ties are resolved does not play an important role in our analysis. It is made here explicit because it is important that $\mathbf{x} \in \mathbb{C}^{n_\text{T}}$ determine the ordering permutation $\sigma(\mathbf{x})$ uniquely.

If $\mathbf{X}$ is a random vector taking value in $\mathbb{C}^{n_\text{T}}$, then its ordering permutation $\sigma(\mathbf{X})$ is a random permutation. Since the number of permutation on $\mathcal{T}$ is $n_\text{T}!$, it follows that, irrespective of the distribution of $\mathbf{X}$, the entropy of $\sigma(\mathbf{X})$ is upper bounded by
$$H\big(\sigma(\mathbf{X})\big) \leq \log n_\text{T}! \ . \tag{34}$$

Given any channel input $\mathbf{X}$ we can thus expand the mutual information $I(\mathbf{X}; \mathbf{Y})$ between the channel terminals as:
$$\begin{aligned} I(\mathbf{X}; \mathbf{Y}) &= I\big(\mathbf{X}, \sigma(\mathbf{X}); \mathbf{Y}\big) \\ &= I\big(\sigma(\mathbf{X}); \mathbf{Y}\big) + I\big(\mathbf{X}; \mathbf{Y}|\sigma(\mathbf{X})\big) \\ &\leq H\big(\sigma(\mathbf{X})\big) + I\big(\mathbf{X}; \mathbf{Y}|\sigma(\mathbf{X})\big) \\ &\leq \log n_\text{T}! + I\big(\mathbf{X}; \mathbf{Y}|\sigma(\mathbf{X})\big) \\ &= \sum_{\tau:\, \Pr[\sigma(\mathbf{X})=\tau]>0} I\big(\mathbf{X}; \mathbf{Y}|\sigma(\mathbf{X})=\tau\big) \Pr[\sigma(\mathbf{X}) = \tau] + \log n_\text{T}! \end{aligned} \tag{35}$$

The proof of the converse will now focus on the terms of the form

$$I(\mathbf{X}; \mathbf{Y} | \sigma(\mathbf{X}) = \tau)$$

where $\tau$ is an arbitrary permutation satisfying

$$\Pr[\sigma(\mathbf{X}) = \tau] > 0.$$

Fix then such a permutation $\tau$ and let

$$\mathcal{E}_\tau = \mathsf{E}\left[\|\mathbf{X}\|^2 | \sigma(\mathbf{X}) = \tau\right]. \tag{36}$$

We will show that corresponding to the set $\mathcal{Z}$ and to the permutation $\tau$ there is a power chain of length $\kappa = \kappa(\mathcal{Z}, \tau)$ such that

$$I(\mathbf{X}; \mathbf{Y} | \sigma(\mathbf{X}) = \tau) \leq \kappa \cdot \log(1 + \log(1 + \mathcal{E}_\tau)) + c \tag{37}$$
$$\leq \kappa^* \cdot \log(1 + \log(1 + \mathcal{E}_\tau)) + c \tag{38}$$

where the constant $c$ depends only on the law of $\mathbb{H}$ and on the permutation $\tau$ but not on the power $\mathcal{E}_\tau$.

Note that once we establish (38), the converse will follow from (36) & (35) and Jensen's inequality by the concavity of the double-logarithmic function. We thus proceed to construct the power chain and to then prove (37).

To simplify the typesetting describing the construction of $\kappa(\mathcal{Z}, \tau)$ we shall use $[\nu]$ for $\tau_\nu$. Thus, conditional on $\sigma(\mathbf{X}) = \tau$ we have that $X([1])$ has the maximal magnitude among all the elements of $\mathbf{X}$, and $X([n_\mathrm{T}])$ has the smallest magnitude.

Let $j_1 = 1$. Assume that we have defined $j_1, \ldots, j_\nu$. We then define $j_{\nu+1}$ as

$$j_{\nu+1} = \min\left\{j_\nu < \ell \leq n_\mathrm{T} : \mathcal{R}_{[\ell]} \setminus \bigcup_{\eta=1}^\nu \mathcal{R}_{[j_\nu]} \neq \emptyset\right\} \tag{39}$$

where the minimum of an empty set should be understood as $\infty$. We then set

$$\kappa = \max\{1 \leq \nu \leq n_\mathrm{T} : j_\nu < \infty\} \tag{40}$$

and define

$$t_\nu = [j_\nu] \qquad \mathcal{B}_\nu = \mathcal{R}_{t_\nu} \setminus \bigcup_{\eta=1}^{\nu-1} \mathcal{R}_{[t_\eta]}, \qquad \nu = 1, \ldots, \kappa. \tag{41}$$

Thus, $t_\nu$ is the next strongest transmitter after $t_{\nu-1}$ that can be heard by some receiver that is uninfluenced by any of the stronger transmitters that are already in the chain. The set $\mathcal{B}_\nu$ is the set of receivers that can hear $t_\nu$ but not any of the stronger transmitters that are in the chain. Note that by (31) we have $\mathcal{R}_{t_1} \neq \emptyset$. In fact, $(t_1, \ldots, t_\kappa)$ is a power chain with respect to $\mathcal{Z}$, so that

$$\kappa \leq \kappa^*. \tag{42}$$

(Recall that $\kappa^*$ is the length of the longest power chain with respect to $\mathcal{Z}$.) Also note that the sets $\{\mathcal{B}_\nu\}$ are disjoint and that by (29) their union is $\mathcal{R}$, i.e., they form a partition of $\mathcal{R}$:

$$\mathcal{R} = \bigcup_{\nu=1}^\kappa \mathcal{B}_\nu \qquad \mathcal{B}_\nu \cap \mathcal{B}_{\nu'} = \emptyset \quad \text{whenever } 1 \leq \nu \neq \nu' \leq \kappa. \tag{43}$$

Finally, we define

$$\mathcal{A}_\nu = \{[j_\nu], \ldots, [j_{\nu+1} - 1]\} \qquad \nu = 1, \ldots, \kappa - 1 \tag{44}$$

and

$$\mathcal{A}_\kappa = \{[j_\kappa], \ldots, [n_\mathrm{T}]\}. \tag{45}$$

The key properties of the constructions of $\kappa(\mathcal{Z}, \tau)$, of the power chain $(t_1, \ldots, t_\kappa)$, of the collection $\{\mathcal{B}_\nu\}$, and of the collection $\{\mathcal{A}_\nu\}$ are as follows. The $\kappa$-tuple $(t_1, \ldots, t_\kappa)$ is a power chain, so that $\kappa \leq \kappa^*$, (42). The collections $\{\mathcal{B}_\nu\}$ and $\{\mathcal{A}_\nu\}$ are partitions of $\mathcal{R}$ and $\mathcal{T}$ respectively. And conditional on $\sigma(\mathbf{X}) = \tau$ the random variables $X(\mathcal{A}_\nu)$ only influence $Y(\cup_{\eta=1}^{\nu} \mathcal{B}_\eta)$; they do not influence any receiver in $\mathcal{R} \setminus \cup_{\eta=1}^{\nu} \mathcal{B}_\eta$. That is, conditional on $\sigma(\mathbf{X}) = \tau$ and on the random variables $X(\mathcal{A}_\nu \cup \cdots \cup \mathcal{A}_\kappa)$, the random variables $X(\mathcal{A}_1 \cup \cdots \cup \mathcal{A}_{\nu-1})$ are independent of the random variables $Y(\mathcal{B}_\nu)$.

Using these properties, we next prove (37). The key will be the following lemma:

**Lemma 3.** *Let $\mathbb{H}$ be a random $n_\mathrm{R} \times n_\mathrm{T}$ complex matrix whose components are all of finite second moment*

$$\mathsf{E}\big[|H(r,t)|^2\big] < \infty, \qquad (r,t) \in \mathcal{R} \times \mathcal{T}$$

*where $\mathcal{R} = \{1, \ldots, n_\mathrm{R}\}$ and $\mathcal{T} = \{1, \ldots, n_\mathrm{T}\}$. Let the set $\mathcal{Z} \subset \mathcal{R} \times \mathcal{T}$ be the set of pairs $(r, t)$ such that $H(r, t)$ is deterministically zero:*

$$H(r,t) = 0 \quad \text{almost surely} \quad \forall (r,t) \in \mathcal{Z}.$$

*Assume that the joint differential entropy of the coordinates that are not in $\mathcal{Z}$ is finite*

$$h\left(H(\mathcal{Z}^\mathrm{c})\right) > -\infty. \tag{46}$$

*Let $t^* \in \mathcal{T}$ be fixed. Assume that Transmitter $t^*$ influences all receivers in the sense that*

$$(r, t^*) \notin \mathcal{Z}, \qquad \forall r \in \mathcal{R}. \tag{47}$$

*Let $\mathbf{X}$ be a random vector taking value in $\mathbb{C}^{n_\mathrm{T}}$ whose component of largest magnitude is almost surely $t^*$:*

$$\max_{t \in \mathcal{T}} |X(t)| = |X(t^*)|, \quad \text{almost surely.} \tag{48}$$

*Assume the average power constraint*

$$\sum_{t \in \mathcal{T}} \mathsf{E}\big[|X(t)|^2\big] \leq \mathcal{E}.$$

*Finally, let $\mathbf{Z}$ take value in $\mathbb{C}^{n_\mathrm{T}}$ according the multivariate Gaussian law $\mathcal{N}_\mathbb{C}(\mathbf{0}, \mathsf{I}_{n_\mathrm{R}})$ and assume that $\mathbb{H}$, $\mathbf{Z}$, and $\mathbf{X}$ are independent.*

*Then there exists some constant $c$, which depends on the law of $\mathbb{H}$ but not on the law of $\mathbf{X}$ or on its power $\mathcal{E}$, such that*

$$I(\mathbf{X}; \mathbb{H}\mathbf{X} + \mathbf{Z}) \leq \log\big(1 + \log(1 + \mathcal{E})\big) + c. \tag{49}$$

*Proof.* Let $\mathbf{Y} = \mathbb{H}\mathbf{X} + \mathbf{Z}$ and let

$$\mathcal{D} = \{\mathbf{x} \in \mathbb{C}^{n_\mathrm{T}} : \max_{1 \leq t \leq n_\mathrm{T}} |x(t)| = |x(t^*)|\} \tag{50}$$

so that (48) can be rewritten as $\Pr[\mathbf{X} \in \mathcal{D}] = 1$.

The proof of the lemma is very similar to the proof of [3, Theorem 4.2]. It too is based on the bound [3, Eq. (333)]:

$$\begin{aligned}
I(\mathbf{X}; \mathbf{Y}) &\leq \log \pi^{n_\mathrm{R}} - \log \Gamma(n_\mathrm{R}) \\
&\quad + \mathsf{E}_\mathbf{X}\left[n_\mathrm{R}\mathsf{E}\left[\log \|\mathbf{Y}\|^2 | \mathbf{X} = \mathbf{x}\right] - h(\mathbf{Y}|\mathbf{X} = \mathbf{x})\right] \\
&\quad + \alpha \left(1 + \log \mathsf{E}[\|\mathbf{Y}\|^2] - \mathsf{E}[\log \|\mathbf{Y}\|^2]\right) \\
&\quad + \log \Gamma(\alpha) - \alpha \log \alpha, \qquad \alpha > 0.
\end{aligned} \tag{51}$$

From this inequality it follows that for inputs $\mathbf{X}$ satisfying (48)

$$\begin{aligned}
I(\mathbf{X}; \mathbf{Y}) &\leq \log \pi^{n_\mathrm{R}} - \log \Gamma(n_\mathrm{R}) \\
&\quad + \sup_{\mathbf{x} \in \mathcal{D}} \left\{n_\mathrm{R}\mathsf{E}\left[\log \|\mathbf{Y}\|^2 | \mathbf{X} = \mathbf{x}\right] - h(\mathbf{Y}|\mathbf{X} = \mathbf{x})\right\} \\
&\quad + \alpha \left(1 + \log \mathsf{E}[\|\mathbf{Y}\|^2] - \mathsf{E}[\log \|\mathbf{Y}\|^2]\right) \\
&\quad + \log \Gamma(\alpha) - \alpha \log \alpha, \qquad \alpha > 0, \quad \Pr[\mathbf{X} \in \mathcal{D}] = 1.
\end{aligned} \tag{52}$$

We now proceed to analyze the various terms in the above. We begin with showing that the supremum, which does not depend on $\mathcal{E}$, is finite

$$\sup_{\mathbf{x} \in \mathcal{D}} \left\{n_\mathrm{R}\mathsf{E}\left[\log \|\mathbf{Y}\|^2 | \mathbf{X} = \mathbf{x}\right] - h(\mathbf{Y}|\mathbf{X} = \mathbf{x})\right\} < \infty. \tag{53}$$

To this end we use Jensen's inequality to obtain:

$$\begin{aligned}
n_\mathrm{R}\mathsf{E}\left[\log \|\mathbf{Y}\|^2 | \mathbf{X} = \mathbf{x}\right] &\leq n_\mathrm{R} \log \mathsf{E}\left[\|\mathbf{Y}\|^2 | \mathbf{X} = \mathbf{x}\right], \quad \mathbf{x} \in \mathbb{C}^{n_\mathrm{T}} \\
&= n_\mathrm{R} \log \left(\mathsf{E}\left[\|\mathbb{H}\mathbf{X}\|^2 | \mathbf{X} = \mathbf{x}\right] + \mathsf{E}\left[\|\mathbf{Z}\|^2\right]\right), \quad \mathbf{x} \in \mathbb{C}^{n_\mathrm{T}} \\
&\leq n_\mathrm{R} \log \left(\mathsf{E}\left[\|\mathbb{H}\|_F^2\right] \cdot \|\mathbf{x}\|^2 + n_\mathrm{R}\right), \quad \mathbf{x} \in \mathbb{C}^{n_\mathrm{T}} \\
&\leq n_\mathrm{R} \log \left(\mathsf{E}\left[\|\mathbb{H}\|_F^2\right] \cdot n_\mathrm{T} \cdot |x(t^*)|^2 + n_\mathrm{R}\right), \quad \mathbf{x} \in \mathcal{D}
\end{aligned} \tag{54}$$

where the second inequality follows from the Cauchy—Schwarz Inequality with $\|\mathbb{H}\|_F^2 = \sum_{r,t} |H(r,t)|^2$ denoting the squared Frobenius norm, and where the last inequality follows by restricting $\mathbf{x}$ to be in the set $\mathcal{D}$ where $\|\mathbf{x}\|^2 \leq n_\mathrm{T}|x(t^*)|^2$.

As to the differential entropy term in (53), we obtain two separate bounds. The first is useful when $\|\mathbf{x}\|^2$ is very small and is otherwise quite crude

$$\begin{aligned}
h(\mathbf{Y}|\mathbf{X} = \mathbf{x}) &= h(\mathbb{H}\mathbf{x} + \mathbf{Z}) \\
&\geq h(\mathbf{Z}) \\
&= n_\mathrm{R} \log(\pi e), \quad \mathbf{x} \in \mathbb{C}^{n_\mathrm{T}}.
\end{aligned} \tag{55}$$

The second is

$$\begin{aligned}
h(\mathbf{Y}|\mathbf{X} = \mathbf{x}) &\geq h(\mathbb{H}\mathbf{x}) \\
&\geq h\left(\mathbb{H}\mathbf{x} \,\big|\, \{H(r,t)\}_{r \in \mathcal{R},\, t \in \mathcal{T} \setminus \{t^*\}}\right) \\
&= h\left(\{H(r', t^*) \cdot x(t^*)\}_{r' \in \mathcal{R}} \,\big|\, \{H(r,t)\}_{r \in \mathcal{R},\, t \in \mathcal{T} \setminus \{t^*\}}\right) \\
&= n_\mathrm{R} \log |x(t^*)|^2 + h\left(\{H(r', t^*)\}_{r' \in \mathcal{R}} \,\big|\, \{H(r,t)\}_{r \in \mathcal{R},\, t \in \mathcal{T} \setminus \{t^*\}}\right) \\
&= n_\mathrm{R} \log |x(t^*)|^2 + h\left(\{H(r', t^*)\}_{r' \in \mathcal{R}} \,\big|\, \{H(r,t)\}_{t \in \mathcal{T} \setminus \{t^*\},\, (r,t) \notin \mathcal{Z}}\right), \\
&\qquad |x(t^*)| > 0.
\end{aligned} \tag{56}$$

Here the first inequality follows because conditioning cannot increase differential entropy; the subsequent equality by expressing $\mathbb{H}\mathbf{x}$ as

$$\mathbb{H}\mathbf{x} = \begin{pmatrix} \sum_{t \in \mathcal{T}} H(1,t)x(t) \\ \vdots \\ \sum_{t \in \mathcal{T}} H(n_\text{R},t)x(t) \end{pmatrix}$$

$$= \begin{pmatrix} H(1,t^*)x(t^*) \\ \vdots \\ H(n_\text{R},t^*)x(t^*) \end{pmatrix} + \begin{pmatrix} \sum_{t \in \mathcal{T} \setminus \{t^*\}} H(1,t)x(t) \\ \vdots \\ \sum_{t \in \mathcal{T} \setminus \{t^*\}} H(n_\text{R},t)x(t) \end{pmatrix},$$

by noting that conditional on $\{H(r,t)\}_{r \in \mathcal{R},\, t \in \mathcal{T} \setminus \{t^*\}}$ the second term on the right is deterministic, and by noting that the addition of a deterministic vector does not affect a vector's differential entropy; the next equality from the behavior of differential entropy under scaling; and the final equality because it is pointless to condition on deterministic random variables. Note that (46) guarantees that the RHS of (56) is finite.

Inequalities (54), (55) and (56) combine to prove (53). The analysis of the other terms in (52) and the choice of $\alpha = \alpha(\mathcal{E})$ in (52) is identical to the analysis in [3, Appendix II]:

$$\log \mathsf{E}\big[\|\mathbf{Y}\|^2\big] = \log\big(\mathsf{E}\big[\|\mathbb{H}\mathbf{X}\|^2\big] + \mathsf{E}\big[\|\mathbf{Z}\|^2\big]\big)$$
$$\leq \log\big(\mathsf{E}\big[\|\mathbb{H}\|^2\big]\mathsf{E}\big[\|\mathbf{X}\|^2\big] + \mathsf{E}\big[\|\mathbf{Z}\|^2\big]\big)$$
$$\leq \log\big(\mathsf{E}\big[\|\mathbb{H}\|_\text{F}^2\big]\mathcal{E} + n_\text{R}\big) \tag{57}$$

$$\mathsf{E}\big[\log \|\mathbf{Y}\|^2\big] = \mathsf{E}\big[\log \|\mathbb{H}\mathbf{X} + \mathbf{Z}\|^2\big] \tag{58}$$
$$\geq \mathsf{E}\big[\log \|\mathbf{Z}\|^2\big] \tag{59}$$

$$\alpha^* = \big(1 + \log \mathsf{E}\big[\|\mathbf{Y}\|^2\big] - \mathsf{E}\big[\log \|\mathbf{Y}\|^2\big]\big)^{-1} \tag{60}$$

where $\alpha^* \downarrow 0$ with the SNR. See [3, Appendix II] for the details. □

**Note 4.** *The Gaussianity of the noise in the above lemma is not crucial. As in [3, Appendix II] the result continues to hold whenever $\mathsf{E}[\|\mathbf{Z}\|^2] < \infty$ is of finite second moment and of finite differential entropy.*

With the aid of this lemma we can now prove (37). We shall upper bound $I(\mathbf{X}; \mathbf{Y}|\sigma(\mathbf{X}) = \tau)$ in $\kappa$ phases. In the first phase we shall upper bound this mutual information by a double-logarithmic term, a constant, and another mutual information term. This latter mutual information term will be upper bounded in the second phase by a double-logarithmic term, a constant, and yet another mutual information term, which is then upper bounded in the third phase. In the final phase, Phase $\kappa$, we upper bound the mutual information by a double-logarithmic term and a constant only, thus terminating the calculation. Since each phase contributes a double-logarithmic term, the $\kappa$ phases contribute together a $\kappa \cdot \log \log \mathcal{E}_\tau$, as required.

The details now follow. In Phase 1 we expand mutual information using the chain rule

$$\begin{aligned}I(\mathbf{X};\mathbf{Y}|\sigma(\mathbf{X})=\tau) &= I\bigl(X(\mathcal{T});Y(\mathcal{R})|\sigma(\mathbf{X})=\tau\bigr)\\ &= I\bigl(X(\mathcal{T});Y(\mathcal{B}_1),Y(\mathcal{B}_1^c)|\sigma(\mathbf{X})=\tau\bigr)\\ &= I\bigl(X(\mathcal{T});Y(\mathcal{B}_1)|\sigma(\mathbf{X})=\tau\bigr)\\ &\quad + I\bigl(X(\mathcal{T});Y(\mathcal{B}_1^c)|Y(\mathcal{B}_1),\sigma(\mathbf{X})=\tau\bigr).\end{aligned} \qquad (61)$$

The first term on the right of the above is easily treated using the lemma, because conditional on $\sigma(\mathbf{X}) = \tau$, the component $X(t_1)$ $(= X([1]))$ is of largest magnitude, and it is heard by all the receivers in $\mathcal{B}_1$. Consequently, we have by Lemma 3

$$I\bigl(X(\mathcal{T});Y(\mathcal{B}_1)|\sigma(\mathbf{X})=\tau\bigr) \le \log\bigl(1+\log(1+\mathcal{E}_\tau)\bigr) + c_1 \qquad (62)$$

where the constant $c_1$ is as in Lemma 3 independent of the SNR.

As for the second term on the RHS of (61) we use the chain rule once again to obtain

$$\begin{aligned}&I\bigl(X(\mathcal{T});Y(\mathcal{B}_1^c)|Y(\mathcal{B}_1),\sigma(\mathbf{X})=\tau\bigr)\\ &\le I\bigl(X(\mathcal{T}),Y(\mathcal{B}_1);Y(\mathcal{B}_1^c)|\sigma(\mathbf{X})=\tau\bigr)\\ &= I\bigl(X(\mathcal{T});Y(\mathcal{B}_1^c)|\sigma(\mathbf{X})=\tau\bigr) + I\bigl(Y(\mathcal{B}_1);Y(\mathcal{B}_1^c)|X(\mathcal{T}),\sigma(\mathbf{X})=\tau\bigr)\\ &= I\bigl(X(\mathcal{A}_1^c);Y(\mathcal{B}_1^c)|\sigma(\mathbf{X})=\tau\bigr) + I\bigl(Y(\mathcal{B}_1);Y(\mathcal{B}_1^c)|X(\mathcal{T}),\sigma(\mathbf{X})=\tau\bigr)\\ &\le I\bigl(X(\mathcal{A}_1^c);Y(\mathcal{B}_1^c)|\sigma(\mathbf{X})=\tau\bigr) + I\left(\{H(r,t)\}_{\substack{t\in\mathcal{T}\\ r\in\mathcal{B}_1}};\{H(r,t)\}_{\substack{t\in\mathcal{T}\\ r\in\mathcal{B}_1^c}}\right)\end{aligned} \qquad (63)$$

Here the last inequality follows by the data processing inequality, and the preceding equality follows because $Y(\mathcal{B}_1^c)$ is conditionally independent of $X(\mathcal{A}_1)$ given $X(\mathcal{A}_1^c)$.

Thus, we have by (61) and (62) that the original mutual information term is upper bounded by a double-logarithmic term, a constant term, and another mutual information term:

$$\begin{aligned}I(\mathbf{X};\mathbf{Y}|\sigma(\mathbf{X})=\tau) &\le \log\bigl(1+\log(1+\mathcal{E}_\tau)\bigr)\\ &\quad + c_1 + I\left(\{H(r,t)\}_{\substack{t\in\mathcal{T}\\ r\in\mathcal{B}_1}};\{H(r,t)\}_{\substack{t\in\mathcal{T}\\ r\in\mathcal{B}_1^c}}\right)\\ &\quad + I\bigl(X(\mathcal{A}_1^c);Y(\mathcal{B}_1^c)|\sigma(\mathbf{X})=\tau\bigr).\end{aligned} \qquad (64)$$

The mutual information term

$$I\bigl(X(\mathcal{A}_1^c);Y(\mathcal{B}_1^c)|\sigma(\mathbf{X})=\tau\bigr)$$

on the RHS of the above is now upper bounded in Phase 2. Notice that this term corresponds to a "smaller" fading channel where the inputs $\mathcal{A}_1$ are immaterial, as are the outputs $\mathcal{B}_1$. In Phase 2 we thus upper bound this term as follows:

$$\begin{aligned}I\bigl(X(\mathcal{A}_1^c);Y(\mathcal{B}_1^c)|\sigma(\mathbf{X})=\tau\bigr) &= I\bigl(X(\mathcal{A}_1^c);Y(\mathcal{B}_1^c\cap\mathcal{B}_2),Y(\mathcal{B}_1^c\cap\mathcal{B}_2^c)|\sigma(\mathbf{X})=\tau\bigr)\\ &= I\bigl(X(\mathcal{A}_1^c);Y(\mathcal{B}_1^c\cap\mathcal{B}_2)|\sigma(\mathbf{X})=\tau\bigr)\\ &\quad + I\bigl(X(\mathcal{A}_1^c);Y(\mathcal{B}_1^c\cap\mathcal{B}_2^c)|Y(\mathcal{B}_1^c\cap\mathcal{B}_2),\sigma(\mathbf{X})=\tau\bigr). \end{aligned} \quad (65)$$

The first term can be bounded using the lemma because $X(t_2)$ is the component of $X(\mathcal{A}_1^c)$ of largest magnitude, and it is heard by all receivers in $\mathcal{B}_1^c \cap \mathcal{B}_2$:

$$I\big(X(\mathcal{A}_1^c); Y(\mathcal{B}_1^c \cap \mathcal{B}_2) \big| \sigma(\mathbf{X}) = \tau\big) \leq \log\big(1 + \log(1 + \mathcal{E}_\tau)\big) + c_2,$$

for some constant $c_2$.

The second term in (65) can be expanded in analogy to (63) to yield

$$I\big(X(\mathcal{A}_1^c); Y(\mathcal{B}_1^c \cap \mathcal{B}_2^c) \big| Y(\mathcal{B}_1^c \cap \mathcal{B}_2), \sigma(\mathbf{X}) = \tau\big)$$
$$\leq I\big(X(\mathcal{A}_1^c), Y(\mathcal{B}_1^c \cap \mathcal{B}_2); Y(\mathcal{B}_1^c \cap \mathcal{B}_2^c) \big| \sigma(\mathbf{X}) = \tau\big)$$
$$= I\big(X(\mathcal{A}_1^c); Y(\mathcal{B}_1^c \cap \mathcal{B}_2^c) \big| \sigma(\mathbf{X}) = \tau\big) + I\big(Y(\mathcal{B}_1^c \cap \mathcal{B}_2); Y(\mathcal{B}_1^c \cap \mathcal{B}_2^c) \big| X(\mathcal{A}_1^c), \sigma(\mathbf{X}) = \tau\big)$$
$$= I\big(X(\mathcal{A}_1^c \cap \mathcal{A}_2^c); Y(\mathcal{B}_1^c \cap \mathcal{B}_2^c) \big| \sigma(\mathbf{X}) = \tau\big) + I\left(\{H(r,t)\}_{\substack{t \in \mathcal{A}_1^c \\ r \in \mathcal{B}_1^c \cap \mathcal{B}_2}}; \{H(r,t)\}_{\substack{t \in \mathcal{A}_1^c \\ r \in \mathcal{B}_1^c \cap \mathcal{B}_2^c}}\right).$$

The mutual information term

$$I\big(X(\mathcal{A}_1^c \cap \mathcal{A}_2^c); Y(\mathcal{B}_1^c \cap \mathcal{B}_2^c) \big| \sigma(\mathbf{X}) = \tau\big)$$

is now upper bounded in Phase 3. This process is continued until the final phase, Phase $\kappa$, when the term

$$I\big(X(\mathcal{A}_1^c \cap \cdots \cap \mathcal{A}_{\kappa-1}^c); Y(\mathcal{B}_1^c \cap \cdots \cap \mathcal{B}_{\kappa-1}^c)\big) = I\big(X(\mathcal{A}_\kappa); Y(\mathcal{B}_\kappa)\big)$$

is upper bounded using the lemma by a double-logarithmic term and a constant without an additional mutual information term. Indeed, the component $X(t_\kappa)$ is of largest magnitude among the terms in $X(\mathcal{A}_\kappa)$ and it influences all the receivers in $\mathcal{B}_\kappa$.

It is thus seen that performing a total of $\kappa$ phases yields the bound (37) and hence, by (42), also (38). The converse now follows from (38) and (35) using Jensen's inequality because the double-logarithmic function is concave and because, in view of (36),

$$\sum_{\tau:\, \Pr[\sigma(\mathbf{X})=\tau]>0} \Pr[\sigma(\mathbf{X}) = \tau] \cdot \mathcal{E}_\tau = \mathsf{E}\big[\|\mathbf{X}\|^2\big]. \tag{66}$$

## 3.2 The Direct Part

To prove the direct part we shall demonstrate that, under the Gaussian marginals assumption, if $(t_1, \ldots, t_\kappa) \in \mathcal{T}^\kappa$ is any power chain with respect to $\mathcal{Z}$ then we can find a distribution on $\mathbf{X}$ under which its components are independent (thus guaranteeing achievability under multiple-access conditions) and such that

$$\varlimsup_{\mathcal{E} \to \infty} \big\{\kappa \cdot \log \log \mathcal{E} - I(\mathbf{X}; \mathbf{Y})\big\} < \infty. \tag{67}$$

The proof that this mutual information is achievable with scalar single-user detectors will be described separately.

Fix then some such power chain $(t_1, \ldots, t_\kappa) \in \mathcal{T}^\kappa$. Consider now a distribution for $\mathbf{X}$ under which the components of $\mathbf{X}$ are independent with marginals that can be described as follows. If some $t \in \mathcal{T}$ is not in $\{t_1, \ldots, t_\kappa\}$, we set $X(t)$ to be deterministically zero

$$t \notin \{t_1, \ldots, t_\kappa\} \Rightarrow (X(t) = 0, \quad \text{a.s.}). \tag{68}$$

As to the other components of $X$, we choose them to be circularly symmetric with squared magnitudes whose logarithms are uniformly distributed on an interval that will be described later:

$$\log |X(t_\nu)|^2 \sim \text{Uniform}\left(\log x_{\min,\nu}^2, \log x_{\max,\nu}^2\right), \qquad \nu = 1, \ldots, \kappa. \tag{69}$$

Here

$$0 < x_{\min,\nu}^2 < x_{\max,\nu}^2 \leq \log \mathcal{E}, \qquad \nu = 1, \ldots, \kappa \tag{70}$$

will be specified later. (See (93) & (94).) Note that with this choice of the marginals,

$$h\left(\log |X(t_\nu)|^2\right) = \log \log \frac{x_{\max,\nu}^2}{x_{\min,\nu}^2}, \qquad \nu = 1, \ldots, \kappa. \tag{71}$$

Since $(t_1, \ldots, t_\kappa)$ is a power chain, it follows that for every $1 \leq \nu \leq \kappa$ we can find a receiver $r_\nu \in \mathcal{R}$ such that

$$r_\nu \in \mathcal{R}_{t_\nu} \setminus \bigcup_{\eta=1}^{\nu-1} \mathcal{R}_{t_\eta}. \tag{72}$$

Thus, Receiver $r_\nu$ can hear Transmitter $t_\nu$

$$(r_\nu, t_\nu) \notin \mathcal{Z} \tag{73}$$

but it is uninfluenced by the transmitters $t_1, \ldots, t_{\nu-1}$

$$(r_\nu, t_\eta) \in \mathcal{Z}, \qquad \eta = 1, \ldots, \nu - 1. \tag{74}$$

It may be influenced by transmitters $t_{\nu+1}, \ldots, t_\kappa$ but those, as we shall see, will be chosen to have powers that are much smaller than the power assigned to Transmitter $t_\nu$.

The mutual information $I(\mathbf{X}; \mathbf{Y})$ can be now lower bounded as follows:

$$\begin{aligned} I(\mathbf{X}; \mathbf{Y}) &= I\left(\{X(t_\nu)\}_{\nu=1}^\kappa; \mathbf{Y}\right) \\ &= \sum_{\nu=1}^\kappa I\left(X(t_\nu); \mathbf{Y} \mid \{X(t_\eta)\}_{\eta=\nu+1}^\kappa\right) \\ &\geq \sum_{\nu=1}^\kappa I\left(X(t_\nu); Y(r_\nu) \mid \{X(t_\eta)\}_{\eta=\nu+1}^\kappa\right). \end{aligned} \tag{75}$$

Here the first equality follows by (68); the second by the chain rule; and the subsequent inequality by restricting the set of observables in each of the terms.

We shall next show that with a judicious choice of the constants

$$\{x_{\min,\nu}\}, \{x_{\max,\nu}\}, \qquad \nu = 1, \ldots, \kappa$$

in (69) we can guarantee that each of the $\kappa$ terms in (75) grows double-logarithmically in the SNR.

The term that is easiest to deal with is the term $I(X(t_\kappa); Y(r_\kappa))$. It corresponds to the mutual information across the terminals of a Ricean fading channel with additive Gaussian noise:

$$Y(r_\kappa) = H(r_\kappa, t_\kappa) X(t_\kappa) + Z(r_\kappa).$$

Indeed, by our choice of $r_\kappa$, non of the transmitters $t_1, \ldots t_{\kappa-1}$ influences it (74), and the other transmitters were chosen deterministically zero (68). This mutual information term can therefore be handled using the results from [3] on the Ricean fading channel.

The other terms, however, are more complicated. Consider the expression

$$I\big(X(t_\nu); Y(r_\nu) \,\big|\, X(t_{\nu+1}) = x(t_{\nu+1}), \ldots, X(t_\kappa) = x(t_\kappa)\big) \tag{76}$$

for some $1 \leq \nu < \kappa$. By (68) and (74) it follows that we can express $Y(r_\nu)$ as:

$$Y(r_\nu) = H(r_\nu, t_\nu) X(t_\nu) + W(r_\nu) \tag{77}$$

where

$$W(r_\nu) = \sum_{\eta=\nu+1}^{\kappa} H(r_\nu, t_\eta) x(t_\eta) + Z(r_\nu), \qquad 1 \leq \nu < \kappa. \tag{78}$$

Since the components of $\mathbb{H}$ are jointly Gaussian, it follows that under the conditioning in (76), the pair $(H(r_\nu, t_\nu), W(r_\nu))$ are jointly Gaussian. They are not, however, independent because the components of $\mathbb{H}$ may be dependent. We thus need to analyze the mutual information across the Ricean fading channel when the additive noise and the multiplicative noise are jointly Gaussian and *dependent*. The following lemma does just that.

**Lemma 5.** *Let the pair of complex random variables $(H, W)$ be jointly Gaussian, and assume that the pair is independent of the complex random variable $X$. Assume that $X$ has a finite second moment and that it is of finite differential entropy. Then,*

$$I(X; HX + W) \geq h(X) - \mathsf{E}\big[\log |X|^2\big] + \mathsf{E}\big[\log |H|^2\big] - \mathsf{E}\left[\log\left(\pi e \left(\sigma_H + \frac{\sigma_W}{|X|}\right)^2\right)\right]. \tag{79}$$

*where $\sigma_W^2, \sigma_H^2 > 0$ are the variances of $W$ and $H$ respectively. Consequently, if the magnitude of $X$ is almost surely larger than the positive constant $x_{\min} > 0$, then*

$$I(X; HX + W) \geq h(X) - \mathsf{E}\big[\log |X|^2\big] + \mathsf{E}\big[\log |H|^2\big]$$
$$- \mathsf{E}\left[\log\left(\pi e \left(\sigma_H + \frac{\sigma_W}{x_{\min}}\right)^2\right)\right], \qquad |X| \geq x_{\min}, \text{ a.s..} \tag{80}$$

*If, additionally, $X$ is circularly symmetric, then*

$$I(X; HX + W) \geq h\big(\log |X|^2\big) + \log \pi + \mathsf{E}\big[\log |H|^2\big]$$
$$- \mathsf{E}\left[\log\left(\pi e \left(\sigma_H + \frac{\sigma_W}{x_{\min}}\right)^2\right)\right], \qquad |X| \geq x_{\min}, \text{ circ. sym..} \tag{81}$$

*Proof.* First note that the assumptions that $X$ has a finite second moment and finite differential entropy guarantee that the logarithm of its magnitude is of finite expectation [3, Lemma 7.7] so that the lemma's claim is meaningful.

The proof proceeds by expressing $I(X; HX + W)$ as

$$I(X; HX + W) = h(HX + W) - h(HX + W | X) \tag{82}$$

and by then bounding the terms on the RHS. We begin with the first:

$$\begin{aligned} h(HX+W) &\geq h(HX+W|H) \\ &\geq h(HX|H) \\ &= h(X) + \mathsf{E}\left[\log |H|^2\right] \end{aligned} \quad (83)$$

where the first inequality follows because conditioning cannot increase differential entropy; the second because conditional on $H$ the random variables $X$ and $W$ are independent; and the subsequent equality from the behavior of differential entropy of *complex* random variables under deterministic scaling.

As to the other term in (82), we note that conditional on $X = x$, the random variable $HX + W$ is Gaussian. Hence,

$$\begin{aligned} h(HX+W|X) &= \mathsf{E}\left[\log |X|^2\right] + h\left(H + \frac{W}{X}\bigg|X\right) \\ &= \mathsf{E}\left[\log |X|^2\right] + \mathsf{E}\left[\log \pi e \cdot \mathsf{Var}\left(H + \frac{W}{X}\bigg|X\right)\right] \\ &\leq \mathsf{E}\left[\log |X|^2\right] + \mathsf{E}\left[\log \pi e \left(\sigma_H + \frac{\sigma_W}{|X|}\right)^2\right] \end{aligned} \quad (84)$$

where $\sigma_H^2$ and $\sigma_W^2$ are the respective variances of $H$ and $W$.

Combining (83) and (84) with (82) yields (79), which combines with the monotonicity of the logarithm function to imply (80). Finally, to obtain (81) we note that if $X$ is circularly symmetric then

$$h(X) - \mathsf{E}\left[\log |X|^2\right] = h\left(\log |X|^2\right) + \log \pi \quad (85)$$

which follows, for example, from [3, Eqs. (320) & (316)]. □

To apply the lemma to the analysis of (76)–(78) we need an estimate on the variance of $W(r_\nu)$. But such an estimate can be readily found using the Cauchy–Schwarz inequality. Under the conditioning in (76), we have from (78)

$$\begin{aligned} \mathsf{Var}(W(r_\nu)) &\leq \mathsf{E}\left[|W(r_\nu)|^2\right] \\ &= \mathsf{E}\left[|Z(r_\nu)|^2\right] + \mathsf{E}\left[\left|\sum_{\eta=\nu+1}^{\kappa} H(r_\nu, t_\eta) x(t_\eta)\right|^2\right] \\ &\leq 1 + \sum_{\eta=\nu+1}^{\kappa} \mathsf{E}\left[|H(r_\nu, t_\eta)|^2\right] \cdot \sum_{\eta=\nu+1}^{\kappa} |x(t_\eta)|^2 \\ &\leq 1 + \mathsf{E}\left[\|\mathbb{H}\|_F^2\right] \cdot (\kappa - \nu) \max_{\nu < \eta \leq \kappa} x_{\max,\eta}^2. \end{aligned} \quad (86)$$

It thus follows from Lemma 5 and from (86) that the mutual information in (76) will satisfy

$$\varlimsup_{\mathcal{E}\to\infty} \left\{\log \log \mathcal{E} - I\big(X(t_\nu); Y(r_\nu) \,\big|\, X(t_{\nu+1}) = x(t_{\nu+1}), \ldots, X(t_\kappa) = x(t_\kappa)\big)\right\} < \infty \quad (87)$$

uniformly over all $\{x(t_\eta), \eta = \nu + 1, \ldots, \kappa\}$ satisfying

$$x_{\min,\eta} \leq |x(t_\eta)| \leq x_{\max,\eta}, \qquad \eta = \nu + 1, \ldots, \kappa \qquad (88)$$

whenever both

$$\lim_{\mathcal{E} \to \infty} \frac{x_{\min,\nu}^2}{1 + \mathsf{E}[\|\mathbb{H}\|_F^2] \cdot (\kappa - \nu) \max_{\nu < \eta \leq \kappa} x_{\max,\eta}^2} = \infty \qquad (89)$$

(so that the last term on the RHS of (81) tends to zero) and

$$\varlimsup_{\mathcal{E} \to \infty} \left\{ \log \log \mathcal{E} - \log \log \frac{x_{\max,\nu}^2}{x_{\min,\nu}^2} \right\} < \infty \qquad (90)$$

(so that by (71) the first term on the RHS of (81) has the right asymptotic growth.) Since (89) and (90) guarantee *uniform* convergence in (87) it follows that they also guarantee that

$$\varlimsup_{\mathcal{E} \to \infty} \left\{ \log \log \mathcal{E} - I\big(X(t_\nu); Y(r_\nu) \,\big|\, X(t_{\nu+1}) \ldots, X(t_\kappa)\big) \right\} < \infty, \qquad \nu = 1, \ldots, \kappa \qquad (91)$$

and hence, by (75), they also guarantee that

$$\varlimsup_{\mathcal{E} \to \infty} \left\{ \kappa \cdot \log \log \mathcal{E} - I(\mathbf{X}; \mathbf{Y}) \right\} < \infty \qquad (92)$$

as we had set out to prove.

To conclude this part of the proof it is thus only required to find choices for $\{x_{\min,\eta}, x_{\max,\eta}\}_{\eta=1}^\kappa$ that will guarantee that both (89) and (90) hold. An example of such a choice is:

$$x_{\max,\eta} = \mathcal{E}^{1/\eta}, \qquad \eta = 1, \ldots, \kappa \qquad (93)$$

$$x_{\min,\eta} = \mathcal{E}^{1/(\eta+1)} \log \mathcal{E}, \qquad \eta = 1, \ldots, \kappa. \qquad (94)$$

Having established the achievability of (22) we now set out to prove that this can also be achieved using $\kappa^*$ scalar single-user detectors. We do so by showing that, with our choice of $\mathbf{X}$, (91) implies that

$$\varlimsup_{\mathcal{E} \to \infty} \left\{ \log \log \mathcal{E} - I\big(X(t_\nu); Y(r_\nu)\big) \right\} < \infty, \qquad \nu = 1, \ldots, \kappa. \qquad (95)$$

Since $I\big(X(t_\nu); Y(r_\nu)\big)$ is achievable with a single-user detector, this will conclude the proof. To prove that (91) implies (95) we will show that

$$\lim_{\mathcal{E} \to \infty} \left\{ I\big(X(t_\nu); Y(r_\nu) \,\big|\, X(t_{\nu+1}), \ldots, X(t_\kappa)\big) - I\big(X(t_\nu); Y(r_\nu)\big) \right\} = 0, \quad \nu = 1, \ldots, \kappa. \qquad (96)$$

To this end we upper bound this difference as:

$$\begin{aligned} I\big(X(t_\nu); Y(r_\nu) \,\big|\, \{X(t_\eta)\}_{\nu < \eta \leq \kappa}\big) &- I\big(X(t_\nu); Y(r_\nu)\big) \\ &= I\big(X(t_\nu); \{X(t_\eta)\}_{\nu < \eta \leq \kappa}, Y(r_\nu)\big) - I\big(X(t_\nu); Y(r_\nu)\big) \\ &= I\big(X(t_\nu); \{X(t_\eta)\}_{\nu < \eta \leq \kappa} \,\big|\, Y(r_\nu)\big) \\ &\leq I\big(X(t_\nu), Y(r_\nu); \{X(t_\eta)\}_{\nu < \eta \leq \kappa}\big) \\ &\leq I\big(X(t_\nu), Y(r_\nu), \{H(r_\nu, t_\eta)\}_{\nu < \eta \leq \kappa}; \{X(t_\eta)\}_{\nu < \eta \leq \kappa}\big) \\ &= I\big(\{X(t_\eta)\}_{\nu < \eta \leq \kappa}; Y(r_\nu) \,\big|\, \{H(r_\nu, t_\eta)\}_{\nu < \eta \leq \kappa}, X(t_\nu)\big). \qquad (97) \end{aligned}$$

Here the first equality follows from the independence of $X(t_\nu)$ and $\{X(t_\eta)\}_{\nu<\eta\leq\kappa}$; the second equality from the chain rule; the subsequent inequality from the chain rule and the nonnegativity of mutual information; the subsequent inequality by adding observations; and the last equality because $\{X(t_\eta)\}_{\nu<\eta\leq\kappa}$ is independent of the pair $(X(t_\nu), \{H(r_\nu, t_\eta)\}_{\nu<\eta\leq\kappa})$. Conditional on

$$\{H(r_\nu, t_\eta)\}_{\nu<\eta\leq\kappa} = \{h(r_\nu, t_\eta)\}_{\nu<\eta\leq\kappa}$$

the distribution of $H(r_\nu, t_\nu)$ is Gaussian with some variance $\epsilon^2 > 0$, which does not depend on the realization of $\{H(r_\nu, t_\eta)\}_{\nu<\eta\leq\kappa}$. (Note that by (26) and (73) $\epsilon^2$ is strictly larger than zero.) Consequently, if we additionally condition on $X(t_\nu) = x(t_\nu)$ we obtain that the mutual information between $Y(r_\nu)$ and $\{X(t_\eta)\}_{\nu<\eta\leq\kappa}$ is the same as the mutual information between $\{X(t_\eta)\}_{\nu<\eta\leq\kappa}$ and $\tilde{Y}$ where

$$\tilde{Y} = \sum_{\eta=\nu+1}^{\kappa} h(r_\nu, t_\eta) X(t_\eta) + W \tag{98}$$

and where

$$W \sim \mathcal{N}_\mathbb{C}(0, 1 + \epsilon^2 |x(t_\nu)|^2)$$

is independent of $\{X(t_\eta)\}_{\nu<\eta\leq\kappa}$. Now

$$\mathsf{E}\left[\left|\sum_{\eta=\nu+1}^{\kappa} H(r_\nu, t_\eta) X(t_\eta)\right|^2 \,\bigg|\, \{H(r_\nu, t_\eta)\}_{\nu<\eta\leq\kappa}, X(t_\nu)\right] = \sum_{\eta=\nu+1}^{\kappa} |H(r_\nu, t_\eta)|^2 \mathsf{E}\big[|X(t_\eta)|^2\big] \tag{99}$$

because the components of $\mathbf{X}$ are independent and circularly symmetric (and hence of zero mean). By the data processing inequality and the Gaussian channel capacity formula we now obtain:

$$I\big(\{X(t_\eta)\}_{\nu<\eta\leq\kappa}; Y(r_\nu) \,\big|\, \{H(r_\nu, t_\eta)\}_{\nu<\eta\leq\kappa}, X(t_\nu)\big)$$
$$\leq \mathsf{E}\left[\log\left(1 + \frac{\sum_{\eta=\nu+1}^{\kappa} |H(r_\nu, t_\eta)|^2 \mathsf{E}[|X(t_\eta)|^2]}{1 + \epsilon^2 \cdot |X(t_\nu)|^2}\right)\right]$$
$$\leq \mathsf{E}\left[\log\left(1 + \frac{\sum_{\eta=\nu+1}^{\kappa} |H(r_\nu, t_\eta)|^2 \mathsf{E}[|X(t_\eta)|^2]}{1 + \epsilon^2 \cdot x_{\min,\nu}^2}\right)\right]$$
$$\leq \log\left(1 + \frac{\sum_{\eta=\nu+1}^{\kappa} \mathsf{E}[|H(r_\nu, t_\eta)|^2] \, \mathsf{E}[|X(t_\eta)|^2]}{1 + \epsilon^2 \cdot x_{\min,\nu}^2}\right)$$
$$\leq \log\left(1 + \frac{\mathsf{E}[\|\mathbb{H}\|_F^2] \cdot (\kappa - \nu) \cdot \max_{\nu<\eta\leq\kappa}\{x_{\max,\eta}^2\}}{1 + \epsilon^2 \cdot x_{\min,\nu}^2}\right)$$
$$\to 0 \tag{100}$$

where the second inequality follows because $|X(t_\nu)| \geq x_{\min,\nu}$ with probability one; the third inequality follows by Jensen's inequality; and the final limiting behavior follows because we have chosen $\{x_{\min,\eta}, x_{\max,\eta}\}$ so that (89) hold.

The upper bound (97) and (100) now combine to prove (96).

# 4  Discussion and Summary

In this paper we considered non-coherent fading networks with additive and multiplicative noises. We have shown that, at high SNR, the capacity of the network grows like an integer multiple of log log SNR. This integer multiple is determined by the location of the deterministic zeroes of the fading matrix. Loosely speaking, this integer can be viewed as the effective number of parallel channels that can be supported by the network, *i.e.*, as the maximal number of point-to-point single-user scalar channels that can be supported by the network in a manner that will allow, with proper power allocation, negligible cross-interference.

It is felt that this integer is an important parameter of the network, but that far more parameters are needed to obtain more precise approximations of the system's throughput.

It has been pointed out to me by Shlomo Shamai that in some broadcast scenarios the fading levels experienced by the different users may be highly correlated so that the assumption that the non-zero components of the fading matrix are of finite joint differential entropy may be violated. Such scenarios can be nevertheless sometimes addressed using our results by noting that in broadcast scenarios the achievable rates are determined by the marginals of the network law [5]. Thus, in some such scenarios one can replace the fading matrix with a fading matrix whose rows are independent, but such that each row is of the same law as that in the original matrix.

# A  On Memory and Feedback at High SNR

In this appendix we show that if the stationary and ergodic fading process $\{\mathbb{H}_k\}$ satisfies (9) and (14) then the difference between the channel's capacity and the capacity of the memoryless channel of IID fading $\{\tilde{\mathbb{H}}_k\}$, where the law of $\tilde{\mathbb{H}}_k$ is identical to the law of $\mathbb{H}_1$, is bounded in the SNR.

This proof is taken almost verbatim from [4] and is included here only for the sake of completeness. We denote the message to be transmitted by $M$, and we assume that the time-$k$ transmitted input $\mathbf{X}_k$ is now a function of $M$ and of the previous outputs $\mathbf{Y}_1^{k-1}$. The proof, as in for example [5, Section 8.12], is based on Fano's inequality and on an upper bound on $n^{-1} \cdot I(M; \mathbf{Y}_1^n)$.

$$\frac{1}{n} I(M; \mathbf{Y}_1^n)$$

$$= \frac{1}{n} \sum_{k=1}^{n} I(M; \mathbf{Y}_k \mid \mathbf{Y}_1^{k-1}) \tag{101}$$

$$= \frac{1}{n} \sum_{k=1}^{n} \left( I(M, \mathbf{Y}_1^{k-1}; \mathbf{Y}_k) - I(\mathbf{Y}_1^{k-1}; \mathbf{Y}_k) \right) \tag{102}$$

$$\leq \frac{1}{n} \sum_{k=1}^{n} I(M, \mathbf{Y}_1^{k-1}; \mathbf{Y}_k) \tag{103}$$

$$\leq \frac{1}{n} \sum_{k=1}^{n} I(M, \mathbf{Y}_1^{k-1}, \mathbb{H}_1^{k-1}; \mathbf{Y}_k) \tag{104}$$

$$= \frac{1}{n} \sum_{k=1}^{n} I(M, \mathbf{Y}_1^{k-1}, \mathbb{H}_1^{k-1}, \mathbf{X}_k; \mathbf{Y}_k) \tag{105}$$

$$= \frac{1}{n} \sum_{k=1}^{n} \left( I(\mathbb{H}_1^{k-1}, \mathbf{X}_k; \mathbf{Y}_k) + I(M, \mathbf{Y}_1^{k-1}; \mathbf{Y}_k \,|\, \mathbb{H}_1^{k-1}, \mathbf{X}_k) \right) \tag{106}$$

$$= \frac{1}{n} \sum_{k=1}^{n} I(\mathbb{H}_1^{k-1}, \mathbf{X}_k; \mathbf{Y}_k) \tag{107}$$

$$= \frac{1}{n} \sum_{k=1}^{n} \left( I(\mathbf{X}_k; \mathbf{Y}_k) + I(\mathbb{H}_1^{k-1}; \mathbf{Y}_k \,|\, \mathbf{X}_k) \right). \tag{108}$$

Here the first two equalities follow from the chain rule; the subsequent inequality from the non-negativity of mutual information; the following inequality from adding random matrices; the subsequent equality follows since $\mathbf{X}_k$ is a deterministic function of $M$ and $\mathbf{Y}_1^{k-1}$; then we have used the chain rule again; (107) follows since

$$I(M, \mathbf{Y}_1^{k-1}; \mathbf{Y}_k \,|\, \mathbb{H}_1^{k-1}, \mathbf{X}_k) = 0; \tag{109}$$

and finally we have used the chain rule once more.

The term $I(\mathbf{X}_k; \mathbf{Y}_k)$ does not depend on the memory in the fading process and is thus identical for $\{\mathbb{H}_k\}$ and for $\{\tilde{\mathbb{H}}_k\}$. As for the other term, we upper bound it as follows:

$$I(\mathbb{H}_1^{k-1}; \mathbf{Y}_k \,|\, \mathbf{X}_k) \leq I(\mathbb{H}_1^{k-1}; \mathbf{Y}_k, \mathbb{H}_k \,|\, \mathbf{X}_k) \tag{110}$$

$$= I(\mathbb{H}_1^{k-1}; \mathbb{H}_k \,|\, \mathbf{X}_k) + I(\mathbb{H}_1^{k-1}; \mathbf{Y}_k \,|\, \mathbf{X}_k, \mathbb{H}_k) \tag{111}$$

$$= I(\mathbb{H}_1^{k-1}; \mathbb{H}_k \,|\, \mathbf{X}_k) \tag{112}$$

$$\leq I(\mathbb{H}_1^{k-1}, \mathbf{X}_k; \mathbb{H}_k) \tag{113}$$

$$= I(\mathbb{H}_k; \mathbb{H}_1^{k-1}) + I(\mathbb{H}_k; \mathbf{X}_k \,|\, \mathbb{H}_1^{k-1}) \tag{114}$$

$$= I(\mathbb{H}_k; \mathbb{H}_1^{k-1}). \tag{115}$$

The feedback capacity of the channel with fading $\{\mathbb{H}_k\}$ can thus exceed the capacity of the memoryless fading channel with equal-marginal fading $\{\tilde{\mathbb{H}}_k\}$ by at most $I(\mathbb{H}_k; \mathbb{H}_1^{k-1})$, which does not depend on the SNR and which, by assumption (14), is finite.

### Acknowledgment

Discussions with Stefan M. Moser and Jatin Thukral are gratefully acknowledged.